\documentclass[reprint,showpacs,showkeys,preprintnumbers,amsmath,amssymb,aps,prb,floatfix]{revtex4-1}

\usepackage[usenames]{color}
\usepackage{graphicx}
\usepackage{dcolumn}
\usepackage{bm}
\usepackage[
  colorlinks=true,
  urlcolor=blue,
  linkcolor=blue,
  citecolor=blue
]{hyperref}

\begin{document}

\title{Self-consistent hybrid functional calculations: Implications
  for structural, electronic, and optical properties of oxide
  semiconductors}

\author{Daniel Fritsch} \email{d.fritsch@bath.ac.uk} \author{Benjamin
  J. Morgan} \author{Aron Walsh} \affiliation{Department of Chemistry,
  University of Bath, Claverton Down, BA2 7AY Bath, UK}

\date{\today}

\begin{abstract}
  The development of new exchange-correlation functionals within
  density functional theory means that increasingly accurate
  information is accessible at moderate computational cost. Recently,
  a newly developed self-consistent hybrid functional has been
  proposed [Skone \emph{et al.} \emph{Phys. Rev. B} \textbf{89} 195112
    (2014)], which allows for a reliable and accurate calculation of
  material properties using a fully \emph{ab initio} procedure. Here,
  we apply this new functional to wurtzite ZnO, rutile SnO$_2$, and
  rocksalt MgO. We present calculated structural, electronic, and
  optical properties, which we compare to results obtained with the
  PBE and PBE0 functionals. For all semiconductors considered here the
  self-consistent hybrid approach gives improved agreement with
  experimental structural data relative to the PBE0 hybrid functional
  for a moderate increase in computational cost, while avoiding the
  empiricism common to conventional hybrid functionals. The electronic
  properties are improved for ZnO and MgO, whereas for SnO$_2$ the
  PBE0 hybrid functional gives best agreement with experimental data.
\end{abstract}



\maketitle

\section{Background}

Metal oxides exhibit many unique structural, electronic, and magnetic
properties, making them useful for a broad range of technological
applications. Metal oxides are exclusively used as transparent
conducting oxides (TCOs) \cite{Minami_SST20_S35}, find applications as
building blocks in artificial multiferroic heterostructures
\cite{Fritsch_PRB82_104117} and as spin-filter devices
\cite{Caffrey_PRB87_024419}, and even include a huge class of
superconducting materials. To develop new materials for specific
applications it is necessary to have a detailed understanding of the
interplay between the chemical composition of different materials,
their structure, and their electronic, optical, or magnetic
properties.

For the development of new functional oxides, computational methods
that allow theoretical predictions of structural and electronic
properties have become an increasingly useful tool. When optical or
electronic properties are under consideration, electronic structure
methods are necessary, with the most popular approach for solids being
density functional theory (DFT). DFT has proven hugely successful in
the calculation of structural properties of condensed matter systems
and the electronic properties of simple metals
\cite{Hasnip_PhilTransA_372_20130270}. The earliest developed
approximate exchange-correlation functionals, however, face
limitations, for example severely underestimating band gaps of
semiconductors and insulators.

Over the last decade several new, more accurate, exchange-correlation
functionals have been proposed. Increased predictive accuracy often
comes with an increased computational cost, and the adoption of these
more accurate functionals has only been made possible through the
continued increase in available computational power. One such more
accurate, and more costly, approach is to use so-called \emph{hybrid}
functionals. These are constructed by mixing a fraction of
Hartree-Fock exact-exchange with the exchange and correlation terms
from some underlying DFT functional. Calculated material properties,
such as lattice parameters and band gaps, however depend on the
precise proportion of Hartree-Fock exact-exchange, $\alpha$. Typical
hybrid functionals treat $\alpha$ as a fixed \emph{empirical}
parameter, chosen by intuition and experimental calibration. A
recently proposed \emph{self-consistent} hybrid functional approach
for condensed systems \cite{Skone_PRB89_195112} avoids this
empiricism, and allows parameter-free hybrid functional calculations
to be performed. In this approach the amount of Hartree-Fock
exact-exchange is identified as the inverse of the dielectric
constant, with this constraint achieved by performing an iterative
sequence of calculations to self-consistency.

Here we apply this new self-consistent hybrid functional to wurtzite
ZnO and rutile SnO$_2$, both materials with potential applications as
TCOs, and MgO, a wide band gap insulator
\cite{Fritsch_APL88_134104}. We examine the implications of the
self-consistent hybrid functional for the structural, electronic, and
optical properties. In the next section we present the theoretical
background, describe the self-consistent hybrid functional, and give
the computational details. We then present results for the structural,
electronic, and optical properties for ZnO, SnO$_2$, and MgO, and
compare these to data calculated using alternative
exchange-correlation functionals and from experiments. The paper
concludes with a summary and an outlook.

\section{Methods}

\subsection{Density functional theory and hybrid functionals}

DFT is a popular and reliable tool to theoretically describe the
electronic structure of both crystalline and molecular systems. DFT
provides a mean-field simplification of the many-body Schr\"odinger
equation. The central variable is the electron density $n(\vec{r}) =
\psi^*(\vec{r})\psi(\vec{r})$, determined from the electronic
wavefunctions $\psi(\vec{r})$, and the Hamiltonian is described as a
functional of the $n(\vec{r})$. Within the generalised Kohn-Sham
scheme the potential is:
\begin{equation}
  v_\mathrm{GKS}(\vec{r}, \vec{r}^\prime) = v_{\mathrm{H}}(\vec{r}) +
  v_\mathrm{xc}(\vec{r},\vec{r}^\prime) + v_\mathrm{ext}(\vec{r}).
\end{equation}
The Hartree potential, $v_{\mathrm{H}}({\bf r})$, and the external
potential, $v_{\mathrm{ext}}({\bf r})$, are in principle known. The
exchange-correlation potential,
$v_\mathrm{xc}(\vec{r},\vec{r}^\prime)$, however, is not, and must be
approximated. Most successful early approximations make use of the
local density approximation and the semilocal generalised gradient
approximation (GGA), for example, in the parametrisation of Perdew,
Burke, and Ernzerhof (PBE) \cite{Perdew_PRL77_3865}. These
approximations already allowed reliable descriptions of structural
properties within the computational resources available at the time,
but lacked accuracy when determining band energies, especially
fundamental band gaps, and $d$ valence band widths of
semiconductors. These properties are particularly important for
reliable calculations of electronic and optical behaviours of
semiconductors.

In recent years so-called \emph{hybrid} functionals have gained in
popularity. In a hybrid functional some proportion of the local
exchange-correlation potential is replaced by Hartree-Fock
exact-exchange terms, giving a better description of electronic
properties. The explicit inclusion of exact-exchange Hartree-Fock
terms make these calculations computationally much more demanding
compared to the earlier GGA calculations, and hybrid functional
calculations have become routine only in recent years. The fraction of
Hartree-Fock exact-exchange admixed in these hybrid functionals,
$\alpha$, is usually justified on experimental or theoretical grounds,
and then fixed for a specific functional. This adds an empirical
parameter and forfeits the \emph{ab initio} nature of the
calculations. One popular choice of $\alpha = 0.25$ is realised in the
PBE0 functional \cite{Adamo_JCP110_6158}.

In this work, we are concerned with full-range hybrid functionals, for
which the generalised nonlocal exchange-correlation potential is
\begin{equation}
  v_\mathrm{xc}(\vec{r},\vec{r}^\prime) = \alpha
  v_\mathrm{x}^\mathrm{ex}(\vec{r},\vec{r}^\prime) + (1-\alpha)
  v_\mathrm{x}(\vec{r}) + v_\mathrm{c}(\vec{r}).
\end{equation}
A common approach is to select $\alpha$ to reproduce the experimental
band gap of solid state systems. Apart from adding an empirical
parameter into the calculations, fitting the band gap of a material
requires reliable experimental data. Moreover this approach does not
guarantee that all electronic properties, e.g., $d$ band widths or
defect levels, are correct \cite{Walsh_PRL100_256401}. Recently, it
has been argued from the screening behaviour of nonmetallic systems
that $\alpha$ can be related to the inverse of the static dielectric
constant
\cite{AlkauskasEtAl_PhysStatSolB2011,MarquesEtAl_PhysRevB2011}
\begin{equation}
  \label{EqAlpha}
  \alpha = \frac{1}{\epsilon_{\infty}},
\end{equation}
which may then be computed in a self-consistent cycle
\cite{Skone_PRB89_195112,Gerosa_PRB91_155201}. This iteration to
self-consistency requires additional computational effort, but removes
the empiricism of previous hybrid functionals and restores the
\emph{ab initio} character of the calculations. The utility of this
approach, however, depends on the accuracy of the resulting predicted
material properties. Here we are interested in the implications for
the structural, electronic, and optical properties of oxide
semidonductors, and consider ZnO, SnO$_2$, and MgO as an illustrative
set of materials.

\subsection{Computational details}

The calculations presented in this work have been performed using the
projector-augmented wave (PAW) method \cite{Bloechl_PRB50_17953}, as
implemented in the Vienna \emph{ab initio} simulation package (VASP
5.4.1)
\cite{Kresse_PRB47_558,Kresse_PRB49_14251,Kresse_CompMatSci6_15}. For
the calculation of structural and electronic properties standard PAW
potentials supplied with VASP were used, with 12 valence electrons for
Zn atom ($4s^{2}3d^{10}$), 14 valence electrons for Sn
($5s^{2}4d^{10}5p^{2}$), 8 valence electrons for Mg ($2p^{6}3s^{2}$),
and six valence electrons for O ($2s^{2}2p^{4}$), respectively. When
calculating dielectric functions we have used the corresponding $GW$
potentials, which give a better description of high-energy unoccupied
states.

To evaluate the performance of the self-consistent hybrid approach, we
have calculated structural and electronic data using three
functionals: GGA in the PBE parametrisation \cite{Perdew_PRL77_3865},
the hybrid functional PBE0 \cite{Adamo_JCP110_6158}, and the
self-consistent hybrid functional \cite{Skone_PRB89_195112}, which we
denote scPBE0.

Structural relaxations were performed for the regular unit cells
within a scalar-relativistic approximation, using dense $k$ point
meshes for Brillouin zone integration ($8\!\times\!8\!\times\!6$ for
wurtzite ZnO, $6\!\times\!6\!\times\!8$ for rutile SnO$_{2}$, and
$10\!\times\!10\!\times\!10$ for rocksalt MgO). For each material, we
performed several fixed-volume calculations, in the cases of ZnO and
SnO$_2$ allowing internal structural parameters to relax until all
forces on ions were smaller than
$0.001\,\text{eV}\,\text{\AA}^{-1}$. Zero-pressure geometries were
determined by then fitting a cubic spline to the total energies with
respect to the unit cell volumes.

To evalutate the self-consistent fraction of Hartree-Fock
exact-exchange, $\alpha$, the dielectric function $\epsilon_{\infty}$
is calculated in an iterative series of full geometry
optimisations. To calculate $\epsilon_{\infty}$, for each of the
ground state structures the static dielectric tensor has been
calculated (including local field effects) from the response to finite
electric fields. For non-cubic systems (ZnO, SnO$_{2}$)
$\epsilon_{\infty}$ was obtained by averaging over the spur of the
static dielectric tensor
$\frac{1}{3}\left(2\epsilon_{\infty}^{\perp}+\epsilon_{\infty}^{\parallel}\right)$. We
have considered $\epsilon_{\infty}$ to be converged when the
difference between two subsequent calculations falls below $\pm0.01$
\cite{Skone_PRB89_195112}.

\section{Results and discussion}

\subsection{Structural properties}

\begin{figure*}
  \includegraphics[width=0.9\textwidth]{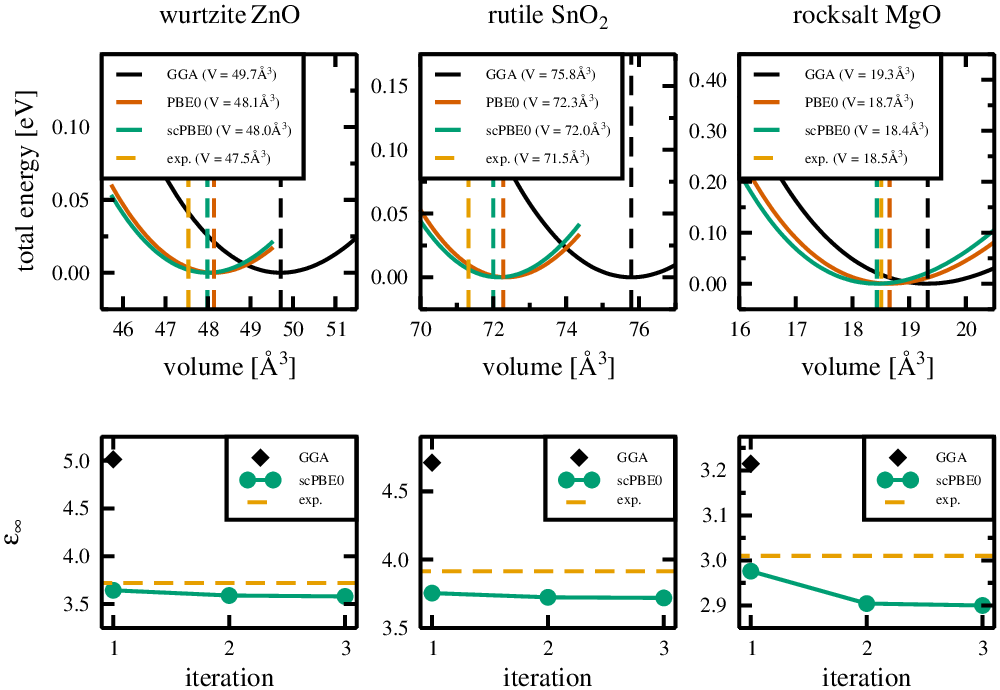}
  \caption{Upper panels: total energy (in eV) with respect to the
    unit cell volume for wurtzite ZnO (left panel), rutile SnO$_{2}$
    (middle panel), and rocksalt MgO (right panel) calculated by means
    of GGA (black), PBE0 (red), and scPBE0 functionals (green),
    respectively. The experimental unit cell volume is depicted by the
    dashed orange line. Lower panels: convergence for the dielectric
    constant $\epsilon_{\infty}$ is obtained after three steps in the
    additional self-consistency cycle.}
  \label{fig1}
\end{figure*}

ZnO crystallises in the hexagonal wurtzite structure of space group
P6$_{3}$mc (No. 186). SnO$_{2}$ crystallises in the tetragonal rutile
structure of space group P4$_{2}$/mnm (No. 136). MgO crystallises in
the cubic rocksalt structure of space group Fm$\bar{3}$m
(No. 225). Each crystal structure was first fully geometry optimised,
as described in the computational details section. The energy / volume
data for the GGA, PBE0, and scPBE0 exchange-correlation potentials are
plotted in the upper panels of Fig.~\ref{fig1}. The GGA functional
significantly overestimates the ground state volume relative to
experimental values for all three materials. This is due to
shortcomings in this simpler early exchange-correlation
potentials. The PBE0 functional adds a fixed proportion of
Hartree-Fock exact-exchange ($\alpha=0.25$), and produces structural
properties in much better agreement with experimental data.

For our scPBE0 calculations, for each material, the static dielectric
constant converged in three iterations (Fig.~\ref{fig1}, lower
panels). Here, computationally the most expensive part is the full
geometry optimisation using the PBE0 functional. Each subsequent step
in the self-consistent loop to determine the amount of Hartree-Fock
exact-exchange starts from optimised crystal structures of the
previous step and reduces the computational costs considerably.

\begin{table}
  \caption{Ground state structural parameters for wurtzite ZnO, rutile
    SnO$_{2}$, and rocksalt MgO obtained with different approximations
    for the exchange-correlation potential in comparison to
    low-temperature experimental data.}
  \label{tab1}
  \begin{tabular}{ccccc}
    ZnO & GGA & PBE0 & scPBE0 & exp. \\
    \hline
    $a$ [\AA] & 3.289 & 3.258 & 3.255 & 3.248 \cite{Reeber_JAP41_5063} \\
    $c$ [\AA] & 5.308 & 5.236 & 5.230 & 5.204 \cite{Reeber_JAP41_5063} \\
    $u$ & 0.381 & 0.381 & 0.381 & 0.382 \cite{Schulz_SSC32_783} \\
    $\varepsilon_{\infty}$ & 5.01 & 3.64 & 3.58 & 3.72 \cite{Heltemes_JAP38_2387} \\
    $\alpha$ & $-$ & 0.25 & 0.28 & $-$ \\
    $E_{\text{gap}}$ [eV] & 0.715 & 3.132 & 3.425 & 3.4449 \cite{Liang_PRL20_59} \\
    
    \hline
    SnO$_{2}$ & GGA & PBE0 & scPBE0 & exp. \\
    \hline
    $a$ [\AA] & 4.834 & 4.757 & 4.752 & 4.737 \cite{Haines_PRB55_11144} \\
    $c$ [\AA] & 3.244 & 3.193 & 3.190 & 3.186 \cite{Haines_PRB55_11144} \\
    $u$ & 0.307 & 0.306 & 0.306 & 0.307 \cite{Haines_PRB55_11144} \\
    $\varepsilon_{\infty}$ & 4.71 & 3.76 & 3.72 & 3.92 \cite{Summitt_JAP39_3762} \\
    $\alpha$ & $-$ & 0.25 & 0.27 & $-$ \\
    $E_{\text{gap}}$ [eV] & 0.609 & 3.591 & 3.827 & 3.596 \cite{Reimann_SSC105_649} \\

    \hline
    MgO & GGA & PBE0 & scPBE0 & exp. \\
    \hline
    $a$ [\AA] & 4.260 & 4.211 & 4.193 & 4.199 \cite{Hazen_AM61_266} \\
    $\varepsilon_{\infty}$ & 3.22 & 2.98 & 2.90 & 3.01 \cite{Jasperse_PR146_526}\\
    $\alpha$ & $-$ & 0.25 & 0.34 & $-$ \\
    $E_{\text{gap}}$ [eV] & 4.408 & 7.220 & 8.322 & 7.833 \cite{Whited_SSC13_1903} \\
  \end{tabular}
\end{table}

Using the self-consistent amount of Hartree-Fock exact-exchange in the
self-consistent hybrid functional yielded structural properties in
slightly better agreement with experimental data (Fig.~\ref{fig1},
upper panels). The improved description of structural properties using
the scPBE0 functional is also evident from the lattice constants $a$
(and $c$), which are given together with those obtained with the other
two functionals and experimental data in Tab.~\ref{tab1}.

\begin{figure}
  \includegraphics[width=0.45\textwidth]{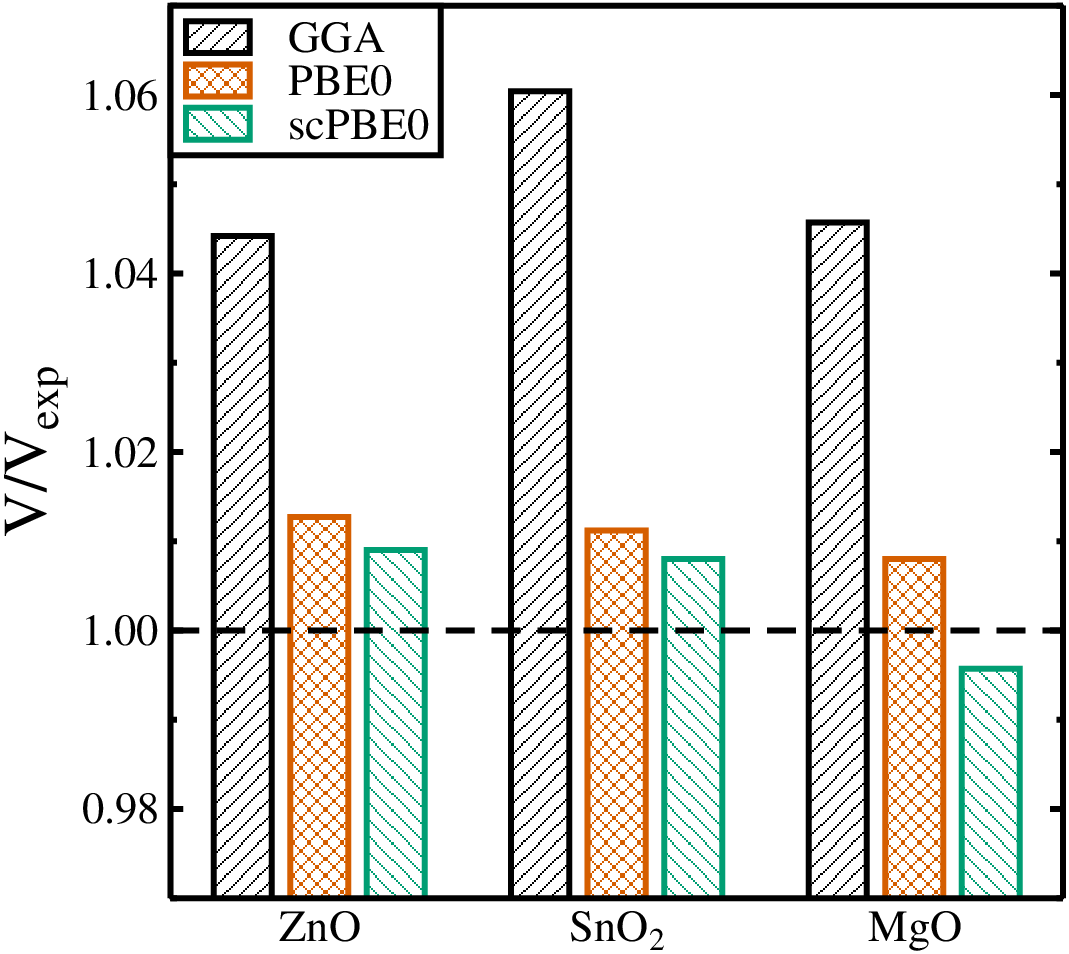}
  \caption{Ground state unit cell volumes $V$ with respect to the
    experimental volume $V_{\mathrm{exp}}$ calculated by means of the
    GGA (black), PBE0 (red), and scPBE0 functionals (green),
    respectively. The experimental volumes correspond to
    $V/V_{\mathrm{exp}}=1$ (dashed horizontal line).}
  \label{fig2}
\end{figure}

The quality of the structural data compared to experiment can also be
seen from Fig.~\ref{fig2} where the coefficients of the different
obtained ground state volumes with respect to the experimental one are
plotted for the three different oxides. Again, the results obtained
with the new self-consistent hybrid functional show closest agreement
with experiment.

\subsection{Electronic and optical properties}

\begin{figure*}
  \includegraphics[width=0.9\textwidth]{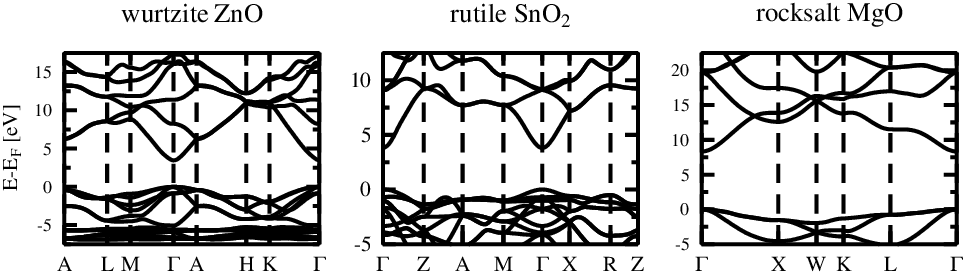}
  \caption{Electronic band structures of wurtzite ZnO (left panel),
    rutile SnO$_{2}$ (middle panel), and rocksalt MgO (right panel),
    calculated with the scPBE0 functional. Energies are in electron
    volt (eV) and the valence band maximum is set to zero.}
  \label{fig3}
\end{figure*}

\begin{figure}
  \includegraphics[width=0.45\textwidth]{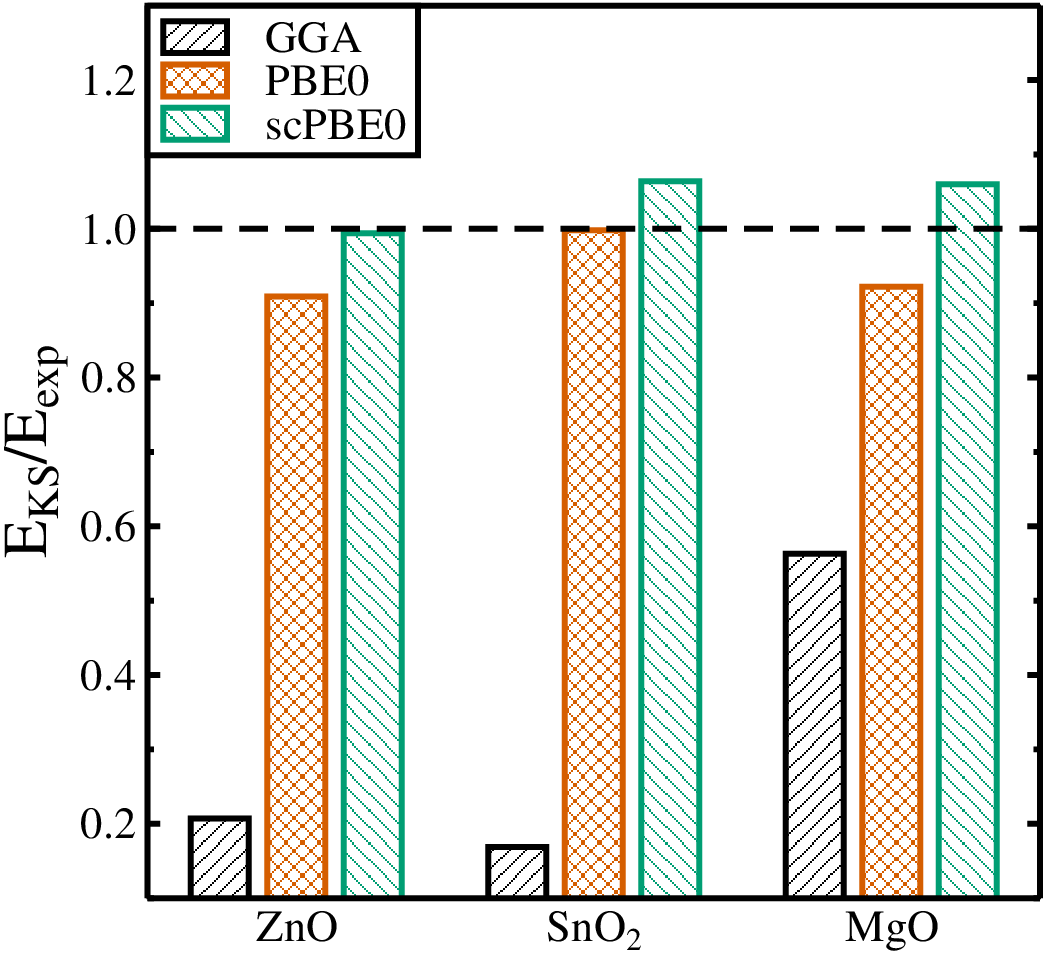}
  \caption{Ground state Kohn-Sham band gaps $E_\mathrm{KS}$ with
    respect to the experimental band gap $E_\mathrm{exp}$ calculated
    by means of the GGA (black), PBE0 (red), and scPBE0 (green),
    respectively. The experimental band gaps correspond to
    $E_\mathrm{KS}/E_\mathrm{exp}=1$ (dashed horizontal line).}
  \label{fig4}
\end{figure}

Fig.~\ref{fig3} shows electronic band structures calculated using
scPBE0 for wurtzite ZnO, rutile SnO$_{2}$, and rocksalt MgO. The
calculated (versus experimental) direct band gaps are
$3.425\,\mathrm{eV}$ ($3.4449\,\mathrm{eV}$ \cite{Liang_PRL20_59}) for
ZnO, $3.827\,\mathrm{eV}$ ($3.596 \,\mathrm{eV}$
\cite{Reimann_SSC105_649}) for SnO$_{2}$, and $8.322\,\mathrm{eV}$
($7.833\,\mathrm{eV}$ \cite{Whited_SSC13_1903}) for MgO, respectively
(Tab.~\ref{tab1}). Fig.~\ref{fig4} shows the GGA, PBE0, and scPBE0
calculated band gaps alongside the experimental values. It can be
seen, that PBE0 calculated band gaps are underestimated compared to
the experimental ones for all three oxides, but being very close for
SnO$_2$. The scPBE0 calculated band gaps are larger than the PBE0
values, thereby improving the results for ZnO and MgO, but worsen the
result for SnO$_2$. In general, the band gaps calculated using the
hybrid functionals (PBE0, scPBE0) are within ten per cent of the
experimental band gaps.

The scPBE0 calculations provide accurate structural properties and
band gaps versus experimental data, and we can therefore be relatively
confident when calculating properties less easily accessible directly
by experiment. We have calculated the real ($\varepsilon_{1}$) and
imaginary ($\varepsilon_{2}$) parts of the dielectric functions via
Fermi's Golden rule summing over transition matrix elements. For these
calculations we used the recommended VASP $GW$ pseudopotentials, and
considerably increased the number of empty bands to ensure converged
results.

\begin{figure*}[t!]
  \includegraphics[width=0.9\textwidth]{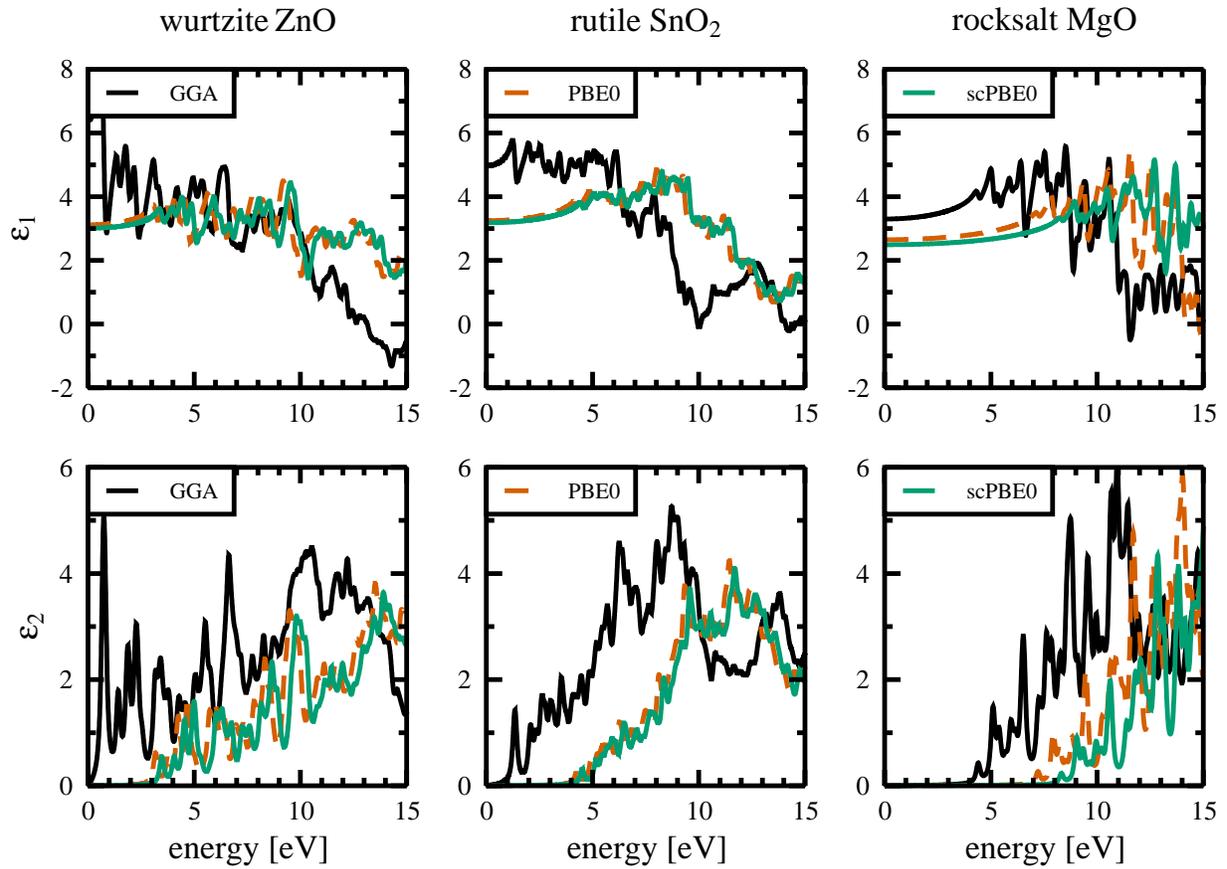}
  \caption{Real $\varepsilon_{1}$ (upper panels) and imaginary
    $\varepsilon_{2}$ (lower panels) parts of the dielectric functions
    calculated by means of GGA (black), PBE0 (dashed red), and scPBE0
    functionals (green), respectively. Dielectric functions are shown
    for wurtzite ZnO (left panels), rutile SnO$_{2}$ (middle panels),
    and rocksalt MgO (right panels).}
  \label{fig5}
\end{figure*}

Fig.~\ref{fig5} shows the real ($\varepsilon_{1}$) and imaginary
($\varepsilon_{2}$) parts of the dielectric functions calculated with
the GGA, PBE0, and scPBE0 functionals. Because the GGA functional
significantly underestimates the band gap, the imaginary parts of the
dielectric functions exhibit an onset, corresponding to the first
allowed direct transition at the fundamental band gap, at lower
energies. The onset energy improves considerably when switching to the
PBE0 hybrid functional, and improves further compared to experiment
when using the self-consistent hybrid functional. For the two hybrid
functionals the overall shape of the real and imaginary parts of the
dielectric functions are very similar in their peak structure, but
differ compared to the pure GGA functional. One reason for this
difference might be the improvements in the $d$ band width and
position when using the hybrid functionals compared to the pure GGA
one. Clarifying this would require a more in-depth comparison of the
different band structures and how their specific features influence
the dielectric functions.

\section{Summary and outlook}

We have presented a theoretical investigation on the application of a
new self-consistent hybrid functional to oxide semiconductors ZnO,
SnO$_{2}$, and MgO. We have presented and compared calculated
structural, electronic, and optical properties of these oxides to
experimental data, and have discussed the implications of using the
new self-consistent hybrid functional. We find that the
self-consistent hybrid functional gives calculated properties with
accuracies as good as or better than the PBE0 hybrid functional. The
additional computational cost due to the self-consistency cycle is
justified by avoiding the empiricism of similar hybrid functionals,
which restores the \emph{ab initio} character of these calculations.

\section{Acknowledgements}

This research has received funding from the European Union's Horizon
2020 research and innovation programme under grant agreement No 641864
(INREP). This work made use of the ARCHER UK National Supercomputing
Service (http://www.archer.ac.uk) via the membership of the UK's HPC
Materials Chemistry Consortium, funded by EPSRC (EP/L000202) and the
Balena HPC facility of the University of Bath. BJM acknowledges
support from the Royal Society (UF130329).

\end{document}